\documentclass{mn2e}
\usepackage{graphicx}

\newif\ifAMStwofonts

\def\mincir{\raise -2.truept\hbox{\rlap{\hbox{$\sim$}}\raise5.truept \hbox{$<$}\ }}
\def\mincireq{\hbox{\raise0.5ex\hbox{$<\lower1.06ex\hbox{$\kern-1.07em{\sim}$}$}}}
\def\magcir{\raise-2.truept\hbox{\rlap{\hbox{$\sim$}}\raise5.truept \hbox{$>$}\ }}

\title{Transparency of the Universe to gamma rays}
\author[De~Angelis, Galanti, Roncadelli]
       {A.~De~Angelis$^{1,2}\thanks{E--mail: deangelis.alessandro@gmail.com}$, G.~Galanti$^{3}$, 
        M.~Roncadelli$^{4}$ \\
        $^1$Dipartimento di Fisica, Universit\`a di Udine, via delle Scienze 208, I-33100 Udine, Italy \\
        $^2$INFN, Sezione di Trieste e Gruppo Collegato di Udine \\ 
        $^3$Dipartimento di Fisica, Universit\`a dell'Insubria, via Valleggio 11, I-22100 Como, Italy \\
        $^4$INFN Pavia, Via A. Bassi 6, I-27100 Pavia, Italy} 

\pubyear{2013}

\begin{document}

\maketitle

\label{firstpage}

\begin{abstract} 

Using the most recent observational data concerning the Extragalactic Background Light and the Radio Background, for a source at a redshift $z_s \leq 3$ we compute the energy $E_0$ of an observed $\gamma$-ray photon in the range $10 \, {\rm GeV} \leq E_0 \leq 10^{13} \, {\rm GeV}$ such that the resulting optical depth $\tau_{\gamma}(E_0,z_s)$ takes the values 1, 2, 3 and 4.6, corresponding to an observed flux dimming of $e^{- 1} \simeq 0.37$, $e^{- 2} \simeq 0.14$, $e^{- 3} \simeq 0.05$ and $e^{- 4.6} \simeq 0.01$, respectively. Below a source distance $D \simeq  8$ kpc we find that $\tau_{\gamma}(E_0,D H_0/c) < 1$ for any value of $E_0$. In the limiting case of a local Universe ($z_s  \simeq 0$) we compare our result with the one derived in 1997 by Coppi and Aharonian. The present achievement is of paramount relevance for the planned ground-based detectors like CTA, HAWC and HiSCORE.

\end{abstract}


\section{Introduction}

Very-high-energy (VHE) astrophysics is on the verge to enter its golden age. Planned ground-based detectors like CTA (Cherenkov Telescope Array)~(Actis et al. 2011), HAWC (High Altitude Water Cherenkov Experiment)~(Sinnis 2005) and HiSCORE (Hundred Square-km Cosmic ORigin Explorer)~(Tluczykont et al. 2011) will probe within the next few years the energy range from $10 \, {\rm GeV}$ up to $10^5 \, {\rm GeV}$ (CTA and HAWC) and even up to $10^9 \, {\rm GeV}$ (HiSCORE) with unprecedented sensitivity. 

Unfortunately, a stumbling block along this exciting avenue is the existence of a soft photon background in the Universe which leads to a strong suppression of the observed flux through the $\gamma \gamma \to e^+ e^-$ pair-production process. The onset of this process depends both on the observed energy and on the source redshift, as it will become apparent later. Specifically, it is mostly due to the Extragalactic Background Light (EBL) -- i.e. to the background in the infrared, visible and ultraviolet region --  in the energy range $10 \, {\rm GeV} - 10^5 \, {\rm GeV}$, to the Cosmic Microwave Background (CMB) in the range $10^5 \, {\rm GeV} - 10^{10} \, {\rm GeV}$ and to the Radio Background (RB) in the range $10^{10} \, {\rm GeV} - 10^{13} \, {\rm GeV}$. Moreover, even though the spectral energy distribution (SED) of the EBL has remained quite uncertain for a long time, a remarkable agreement among the various EBL models has been reached recently, and also the RB has been measured in 2008 with considerably better precision than before.

All this prompts us to carefully evaluate the optical depth -- which quantifies the photon absorption due to the above pair-production process -- in the energy range $10 \, {\rm GeV} - 10^{13} \, {\rm GeV}$ for a source redshift up to $z_s = 3$. As a particular case, we derive the photon mean free path for $\gamma \gamma \to e^+ e^-$ as a function of energy in the local Universe ($z_s \simeq 0$) in order to compare it with that obtained by Coppi and Aharonian (CA) in 1997~(Coppi \& Aharonian 1997). A less detailed but similar result was independently derived at the same time by Phrotheroe and Biermann~(Phrotheroe \& Biermann 1997).

\section{Evaluation of the pair-production cross-section}

We start by recalling that the photon survival probability $P_{\gamma \to \gamma} (E_0,z_s)$ is currently parameterized as 
\begin{equation}
P_{\gamma \to \gamma} (E_0,z_s) = e^{- \tau_{\gamma}(E_0,z_s)}~, 
\label{probcp}
\end{equation}
where $E_0$ is the observed energy and $\tau_{\gamma}(E_0,z_s)$ is the optical depth that quantifies the dimming of the source at redshift $z_s$. Clearly ${\tau}_{\gamma}(E_0,z_s)$ increases with $z_s$, since a greater source distance implies a larger probability for a photon to disappear. Apart from atmospheric effects, one typically has $\tau_{\gamma}(E_0,z_s) < 1$ for $z_s$ not too large, in which case the Universe is optically thin all the way out to the source. But depending on $E_0$ it can happen that $\tau_{\gamma}(E_0,z_s) > 1$, and so at some point the Universe becomes optically thick along the line of sight to the source. The value $z_h$ such that $\tau_{\gamma}(E_0,z_h) = 1$ defines the $\gamma$-{\it ray horizon} for a given $E_0$, and sources beyond the horizon (namely with $z_s > z_h$) tend to become progressively invisible as $z_s$ further increases past $z_h$. 

Whenever dust effects can be neglected, photon depletion arises solely when hard photons of energy $E$ scatter off soft background photons of energy $\epsilon$ permeating the Universe, which gives rise to hard photon absorption. Let us proceed to quantify this issue.

Regarding $E$ as an independent variable, the process is kinematically allowed for
\begin{equation} 
\label{eq.sez.urto01012011}
\epsilon > {\epsilon}_{\rm thr}(E,\varphi) \equiv \frac{2 \, m_e^2 \, c^4}{ E \left(1-\cos \varphi \right)}~,
\end{equation}
where $\varphi$ denotes the scattering angle and $m_e$ is the electron mass. Note that $E$ and $\epsilon$ change along the line of sight in proportion of $1 + z$ because of the cosmic expansion. The corresponding Breit-Wheeler cross-section is~(Breit \& Wheeler 1934; Heitler 1960)
\begin{eqnarray} 
\sigma_{\gamma \gamma}(E,\epsilon,\varphi)  &=& \frac{2\pi\alpha^2}{3m_e^2}  W(\beta) \\
                                                        & \simeq& 1.25 \cdot 10^{-25} \,  W(\beta) \,  {\rm cm}^2~, \label{eq.sez.urto} \nonumber
\end{eqnarray}
with
\[ W(\beta) = \left(1-\beta^2 \right) \left[2 \beta \left( \beta^2 -2 \right) 
+ \left( 3 - \beta^4 \right) \, {\rm ln} \left( \frac{1+\beta}{1-\beta} \right) \right] \, . \]
The cross-section depends on $E$, $\epsilon$ and $\varphi$ only through the  speed $\beta$ -- in natural units -- of the electron and of the positron in the center-of-mass
\begin{equation} 
\label{eq.sez.urto01012011q}
\beta(E,\epsilon,\varphi) \equiv \left[ 1 - \frac{2 \, m_e^2 \, c^4}{E \epsilon \left(1-\cos \varphi \right)} \right]^{1/2}~, 
\end{equation}
and Eq. (\ref{eq.sez.urto01012011}) implies that the process is kinematically allowed for ${\beta}^2 > 0$. The cross-section $\sigma_{\gamma \gamma}(E,\epsilon,\varphi)$ reaches its maximum ${\sigma}_{\gamma \gamma}^{\rm max} \simeq 1.70 \cdot 10^{- 25} \, {\rm cm}^2$ for $\beta \simeq 0.70$. Assuming head-on collisions for definiteness ($\varphi = \pi$), it follows that $\sigma_{\gamma \gamma}(E,\epsilon,\pi)$ gets maximized for the background photon energy 
\begin{equation} 
\label{eq.sez.urto-0}
\epsilon (E) \simeq \left(\frac{500 \, {\rm GeV}}{E} \right) \, {\rm eV}~,
\end{equation}
where $E$ and $\epsilon$ correspond to the same redshift. 
 For an isotropic background of photons, the cross-section is maximized for background photons of energy (Gould \& Schreder 1967)
 \begin{equation} 
 \label{eq.sez.urto-1}
\epsilon (E) \simeq \left(\frac{900 \, {\rm GeV}}{E} \right) \, {\rm eV}~.
\end{equation}
 
 Explicitly, the situation can be summarized as follows.
\begin{itemize}

\item For $10 \, {\rm GeV} \leq E < 10^5 \, {\rm GeV}$ the EBL plays the leading role. In particular, 
for $E \sim 10 \,{\rm GeV}$ $\sigma_{\gamma \gamma}(E,\epsilon)$ -- integrated over an isotropic distribution of background photons -- is maximal for $\epsilon \sim 90 \, {\rm eV}$, corresponding to far-ultraviolet soft photons, whereas for $E \sim 10^5 \,{\rm GeV}$ $\sigma_{\gamma \gamma}(E,\epsilon)$ is maximal for $\epsilon \sim 9 \cdot 10^{- 3} \, {\rm eV}$, corresponding to soft photons in the far-infrared.

\item For $10^5 \, {\rm GeV} \leq E < 10^{10} \, {\rm GeV}$ the interaction with the CMB becomes dominant.

\item For $E \geq 10^{10} \, {\rm GeV}$ the main source of opacity of the Universe is the RB.

\end{itemize}

\section{Evaluation of the optical depth}

Within the standard $\Lambda$CDM cosmological model ${\tau}_{\gamma}(E_0,z_s)$ arises by first convolving the spectral number density $n_{\gamma}({\epsilon}(z), z)$ of background photons at a generic redshift $z$ with ${\sigma}_{\gamma \gamma} (E(z), {\epsilon}(z), \varphi)$ for fixed values of $z$, $\varphi$ and ${\epsilon}(z)$, and next integrating over all these variables~(Gould \& Schreder 1967;  Fazio \& Stecker 1970). Hence, we have
\begin{eqnarray}
\label{eq:tau}
\tau_{\gamma}(E_0, z_s) = \int_0^{z_s} {\rm d} z ~ \frac{{\rm d} l(z)}{{\rm d} z} \, \int_{-1}^1 {\rm d}({\cos \varphi}) ~ \frac{1- \cos \varphi}{2} \ 
\times \\
\nonumber
\times  \, \int_{\epsilon_{\rm thr}(E(z) ,\varphi)}^\infty  {\rm d} \epsilon(z) \, n_{\gamma}(\epsilon(z), z) \,  
\sigma_{\gamma \gamma} \bigl( E(z), \epsilon(z), \varphi \bigr)~, \ \ 
\end{eqnarray}
where the distance travelled by a photon per unit redshift at redshift $z$ is given by
\begin{equation}
\label{lungh}
\frac{{\rm d} l(z)}{{\rm d} z} = \frac{c}{H_0} \frac{1}{\left(1 + z \right) \left[ {\Omega}_{\Lambda} + {\Omega}_M \left(1 + z \right)^3 \right]^{1/2}}~,
\end{equation}
with Hubble constant $H_0 \simeq 70 \, {\rm km} \, {\rm s}^{-1} \, {\rm Mpc}^{-1}$, while ${\Omega}_{\Lambda} \simeq 0.7$ and ${\Omega}_M \simeq 0.3$ represent the average cosmic density of matter and dark energy, respectively, in units of the critical density ${\rho}_{\rm cr} \simeq 0.97 \cdot 10^{- 29} \, 
{\rm g} \, {\rm cm}^{ - 3}$. 

Once $n_{\gamma}(\epsilon(z), z) $ is known, $\tau_{\gamma}(E_0, z)$ can be computed exactly; generally the integration over $\epsilon (z)$ in Eq. (\ref{eq:tau}) must  be performed numerically.   

Finally, in order to get a feeling about the considered physical situation, it looks suitable to discard cosmological effects, which evidently makes sense only for $z_s$ small enough. Accordingly, $z_s$ is best expressed in terms of the source distance $D = c z_s /H_0$, and the optical depth becomes\footnote{Since in this case the energy is independent of the source distance, we simply write the observed energy as $E$ instead of $E_0$.} 
\begin{equation}
\label{lungh26122010S}
\tau_{\gamma}  = \frac{D}{{\lambda}_{\gamma}(E)}~,
\end{equation}
where ${\lambda}_{\gamma}(E) = D/\tau_\gamma(E, DH_0/c)$ is the photon mean free path for $\gamma \gamma \to e^+ e^-$ referring to the present cosmic epoch. As a consequence, Eq. (\ref{probcp}) becomes
\begin{equation} 
\label{a012122010W}
P_{\gamma \to \gamma} (E,D) = e^{- D/{\lambda}_{\gamma}(E)}~.
\end{equation}

\section{Soft photon background}

Our main goal is at this point the determination of $n_{\gamma}(\epsilon(z), z) $. For the sake of clarity, we consider separately the EBL, the CMB, and the RB.

The EBL density $n_{\gamma}(\epsilon(z), z) $ is in principle affected by large uncertainties arising mainly from foreground contamination produced by zodiacal light which is various orders of magnitude larger than the EBL itself~(Hauser \& Dwek 2001). Below, we sketch schematically the different approaches that have been pursued, without any pretension of completeness.

\begin{itemize}

\item {\it Forward evolution} -- This is the most ambitious approach, since it starts from first principles, namely from semi-analytic models of galaxy formation in order to predict the time evolution of the galaxy luminosity function~(Primack et al. 2001; Primack, Bullock \& Sommerville 2005; Gilmore et al. 2009; Gilmore et al. 2012). 

\item {\it Backward evolution} -- This begins from observations of the present galaxy population and extrapolates the galaxy luminosity function backward in time. Among others, this strategy has been followed by Stecker, Malkan and Scully~(Stecker, Malkan \& Scully, 2006) -- whose result has unfortunately been ruled out by the measurements by {\it Fermi}/LAT~(Abdo et al. 2010) -- and by Franceschini, Rodighiero and Vaccari (FRV)~(Franceschini, Rodighiero \& Vaccari 2008).

\item {\it Inferred evolution} -- This approach models the EBL by using quantities like the star formation rate, the initial mass function and the dust extinction as inferred from observations~(Kneiske, Mannheim \& Hartmann 2002; Kneiske et al. 2004; Finke, Razzaque \& Dermer 2010).

\item {\it Observed evolution} -- This method developed by Dominguez and collaborators (D) relies on observations by using a very rich sample of galaxies extending over the redshift range 
$0 \leq z_s  \leq 2$~(Dominguez et al. 2011).

\item {\it Compared observations} -- This technique has been implemented in two different ways. One consists in comparing observations of the EBL itself with blazar observations with Imaging Atmospheric Cherenkov Telescopes (IACTs) and deducing the EBL level from the VHE photon dimming~(Schr\"odter 2005; Aharonian et al. 2006a; Mazin \& Raue 2007; Mazin \& G\"obel 2007; Finke \& Razzaque 2009). The other starts from some $\gamma$-ray observations of a given blazar below $100 \, {\rm GeV}$ where EBL absorption is negligible -- typically using {\it Fermi}/LAT data -- and infers the EBL level by comparing the IACT observations of the same blazar with the source spectrum as extrapolated from  former observations~(Orr, Krennrich \& Dwek 2011) (but see also~Costamante 2012). In the latter case the main assumption is that the emission mechanism is presumed to be determined with great accuracy. In either case, the crucial unstated assumption is that photon propagation in the VHE band is governed by conventional physics.

\item {\it Empirical determination} -- A newer method of determining the EBL, made possible by recent extensive deep galaxy surveys, is to use the observed luminosity densities at different wavelengths together with observational error bars to directly determine the EBL and opacity without the need of any theoretical assumption (Stecker, Malkan \& Scully 2012).

\item {\it Minimal EBL model} -- Its aim is to provide a strict lower limit on the EBL level. It relies on the same strategy underlying the {\it inferred evolution}, but with the parameters tuned in such a way to reproduce the EBL measurements from galaxy counts~(Kneiske \& Dole 2010).

\end{itemize}

Quite remarkably, all methods -- apart obviously from the last one -- yield basically the same results in the redshift range where they overlap, so that at variance with the time when the CA analysis was done (1997) nowadays the SED of the EBL is fixed to a very good extent. We will employ here the FRV model~(Franceschini, Rodighiero \& Vaccari 2008) but we shall check our results using the D model~(Dominguez et al. 2011). As far as the CMB is concerned we take the standard temperature value $T = 2.73 \, {\rm K}$; due to the large density of CMB photons, this background corresponds to the
minimum mean-free-path (Gould \& Schreder 1967; Stecker 1971). Finally, the most recent available data for the RB are employed~(Gervasi et al. 2008), with a low-frequency cutoff taken at 2 MHz.

\section{Results}

\begin{figure*}
\centering
\includegraphics[width=2.0\columnwidth]{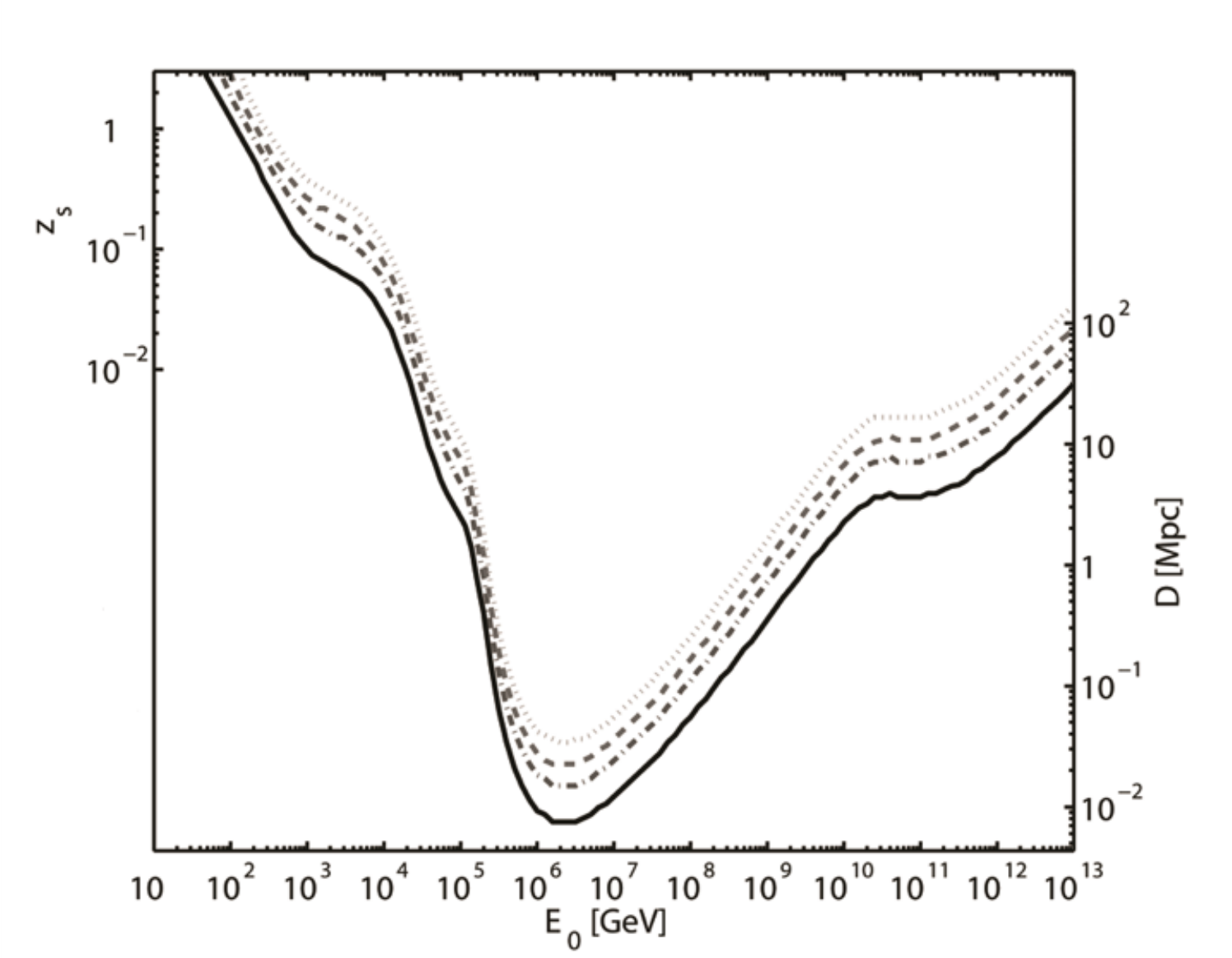}
\caption{\label{fig1ultima} 
Source redshifts $z_s$ at which the optical depth takes fixed values as a function of the observed hard photon energy $E_0$; the $y$-scale on the right side shows the  distance in Mpc for nearby sources. The curves from bottom to top correspond to a photon survival probability of $e^{- 1} \simeq 0.37$ (the horizon), $e^{- 2} \simeq 0.14$, $e^{- 3}  \simeq 0.05$  and $e^{- 4.6} \simeq 0.01$. For $D \simeq 8$ kpc the photon survival probability is larger than $0.37$ for any value of $E_0$.}    
\end{figure*}

Taking into account the EBL, the CMB, and the 
RB, we directly evaluate $\tau_{\gamma}(E_0, z_s)$ over the energy range $10 \, {\rm GeV} \leq E_0 \leq 10^{13} \, {\rm GeV}$ and within the source redshift interval $10^{- 3} \leq z_s \leq 3$, which is the range where the FRV model yields the contribution of the EBL to $\tau_{\gamma}(E_0, z_s)$ (the D model is restricted to $10^{- 2} \leq z_s \leq 2$). We linearly extrapolate $\tau_{\gamma}(E_0, D H_0/c)$ down to $D \simeq 4 \, {\rm kpc}$. Such an extrapolation looks quite reliable not only since $\tau_{\gamma}(E_0, D H_0/c)$ behaves linearly already in  the range $4 \, {\rm Mpc} \leq D \leq 43 \, {\rm Mpc}$, but mainly because at such low distances Eq. (\ref{lungh26122010S}) indeed implies $\tau_{\gamma}(E_0, D H_0/c) \propto D H_0/c$. Moreover, we stress that the extrapolation in question does not practically affect our result. As it is evident from Fig. \ref{fig1ultima}, for $D < 4 \, {\rm Mpc}$ the energies $E_0$ for which $\tau_{\gamma}(E_0, D H_0/c)$ takes our prescribed values exceed $\sim 10^5 \, {\rm GeV}$, which means that our result for 
$D \simeq 4 \, {\rm Mpc}$ is dominated by the CMB and the RB rather than by the EBL. Although we have computed $\tau_{\gamma}(E_0, z_s)$ using the FRV model, we have checked that it basically remains unaffected by employing the D model in the redshift range where they overlap. This is our main result, which is plotted in Fig. \ref{fig1ultima}, where the solid line corresponds to $\tau_{\gamma}(E_0, z_s) = 1$, the dot-dashed line corresponds to $\tau_{\gamma}(E_0, z_s) = 2$, the dashed line corresponds to $\tau_{\gamma}(E_0, z_s) = 3$ and the dotted line corresponds to $\tau_{\gamma}(E_0, z_s) = 4.6$, which give rise to an observed flux dimming of about $0.37$, $0.14$, $0.05$ and $0.01$, respectively. For $D \leq 8 \, {\rm kpc}$ it turns out that $\tau_{\gamma}(E_0, D H_0/c) < 1$ for any value of $E_0$.

In order to compare our finding with the CA result, we disregard cosmological effects thereby computing the $\gamma \gamma \to e^+ e^-$ mean free path $\lambda_{\gamma} (E)$ in the local Universe ($z_s \simeq 0$) by using Eq. (\ref{lungh26122010S}) with $\tau_{\gamma}(E_0, z_s)$ evaluated by means of Eq. (\ref{eq:tau}) with $D \simeq 4 \, 
{\rm Mpc}$ (formally $z_s \simeq 10^{- 6}$).

\begin{figure}
\centering
\includegraphics[width=\columnwidth]{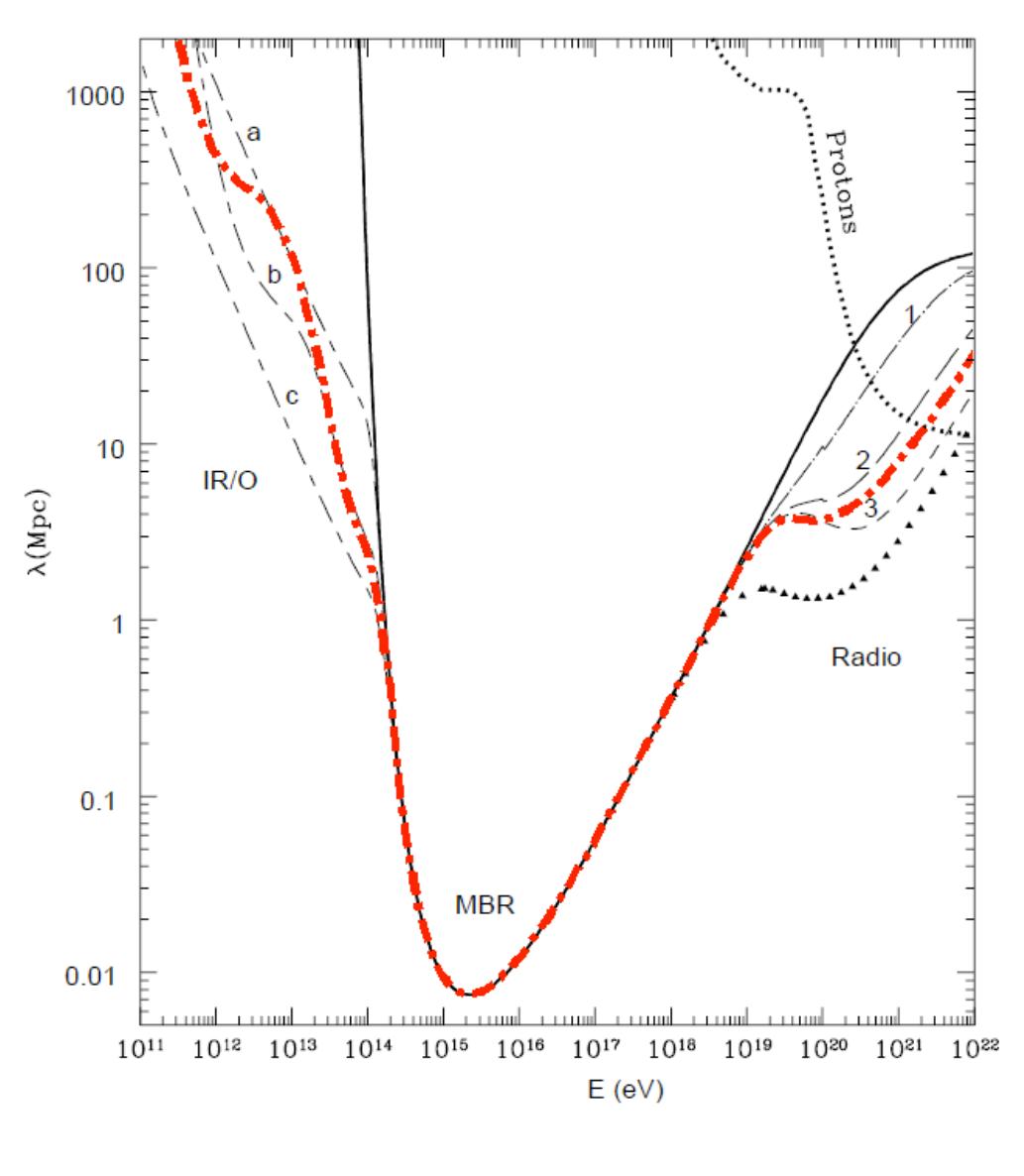}
\caption{\label{fig2ultima} 
Comparison of the mean free path $\lambda_{\gamma}$ for $\gamma \gamma \to e^+ e^-$ as derived in the text using a low-frequency cutoff at 2 MHz (red dot-dashed line) with the one obtained by CA in the local Universe ($z_s = 0$) as a function of the observed energy $E$. The CA black lines labelled by $a$, $b$, $c$ represent three different EBL models, the black solid line corresponds to the CMB, the black lines labelled by $1$, $2$, $3$ represent a model of the RB with low-frequency cutoff at 5, 2 and 1 MHz, respectively, and the black triangles correspond to the RB under the assumption that it is completely extragalactic.} 
\end{figure}

To facilitate the comparison between our and the CA results we have superposed in Fig. \ref{fig2ultima} the behavior of $\lambda_{\gamma} (E)$ as found with the above procedure -- represented by red dot-dashed line -- over Fig. 1 of CA, where the black lines represent the similar results derived by them (see captions). There is of course little surprise that at $z_s \simeq 0$ our result is barely in agreement with that of CA at $z_s = 0$. Still, it should be appreciated that the improved behaviour of $\lambda_{\gamma} (E)$ found here is in disagreement with the one arising from a {\it single} choice of the EBL model in the CA analysis. In addition, while the CMB contribution is obviously identical in both cases, as far as the RB is concerned our curve -- which corresponds to a low-frequency cutoff of 2 MHz -- lies between those corresponding to the low-frequency cutoffs of 2 MHz  and 1 MHz, respectively, of the CA result.

\section{Conclusions}

We have quantified by means of the optical depth the photon absorption caused by the pair-production process in the observed energy range $10 \, {\rm GeV} \leq E_0 \leq 10^{13} \, {\rm GeV}$ and for a source redshift up to $z_s = 3$. We have found that depending on $E_0$ the absorption can be quite large, and becomes dramatic around $10^6 \, {\rm GeV}$. However, for a source distance $D \leq 8 \, {\rm kpc}$ the absorption is irrelevant for any value of $E_0$. 

As it is clear from Fig. \ref{fig1ultima}, our conclusion is of great importance for the planned VHE detectors like CTA, HAWC and HiSCORE. 

Moreover, we have specialized our analysis to the local Universe ($z_s \simeq 0$) where the optical depth is more conveniently replaced by the mean free path, and we have compared it with the same quantity as evaluated by Coppi and Aharonian in 1997.

It has been realized that the blazars observed so far by IACTs give rise to the {\it pair-production anomaly}. The H.E.S.S. collaboration~(Aharonian et al. 2006b) first observed that the SED from the  blazars H 2356-309 and 1ES 1101-232, at redshifts $ z_s = 0.165$ and $z_s = 0.186,$  respectively, could be explained  by a very low EBL level; subsequently, the MAGIC collaboration~(MAGIC Collaboration 2008) have reinforced this evidence with the data from the AGN 3C279 at $z_s = 0.54$. Later, De Angelis et al.~(De Angelis et al. 2009) have observed this effect in the spectral indices at VHE of a sample of AGN at  $z_s > 0.2$. Recently a statistical analysis of the SED from all blazars observed at VHE indicates a level of the EBL {lower} than that predicted even by the {\it minimal EBL model}; the indication  remains at a confidence level between 2.6 and 4.3 depending on the adopted EBL model~(Horns \& Meyer 2012; Meyer, Horns \& Raue 2012). Among many other things, we find it very interesting to see whether this effect persists also at energies much higher than those presently explored by IACTs, and our result looks essential as a benchmark for comparison.

In a forthcoming paper an analysis along the same lines of this work will be carried out taking Axion-like particles (ALPs) into account, since photon-ALP oscillations tend to drastically reduce photon absorption effects of the kind considered here, thereby considerably enlarging the $\gamma$-ray horizon at the VHEs that CTA, HAWC and HiSCORE will be able to probe (see De Angelis, Galanti \& Roncadelli 2011 and references therein). Moreover, it has very recently been pointed out that such a mechanism would explain the pair-production anomaly in a natural fashion and works for values of the photon-ALP coupling in the reach of the planned upgrade of the ALPS experiment at DESY~(Meyer, Horns \& Raue 2013).

Another possible mechanism to explain the apparent reduction of the cosmic gamma-opacity contemplates an additional contribution of secondary gamma rays arising from  interactions along the line of sight of high energy cosmic rays produced by the source~(Essey et al. 2011).

\section*{Acknowledgments}

G. G. thanks Giorgio Sironi for suggestions and M. R. acknowledges the INFN grant FA51.

\newcommand{\BY}[1]{{#1},}
\newcommand{\IN}[4]{{#1} \textbf{#2} (#3) #4}
\newcommand{\TITLE}[1]{ #1 }

\end{document}